\documentclass[manuscript,screen,review=false, acmsmall]{acmart}

\AtBeginDocument{%
  \providecommand\BibTeX{{%
    \normalfont B\kern-0.5em{\scshape i\kern-0.25em b}\kern-0.8em\TeX}}}

\setcopyright{acmcopyright}
\copyrightyear{2023}
\acmYear{2023}
\acmDOI{XXXXXXX.XXXXXXX}

\begin{document}

\title{UNDERSTANDING RUG PULLS: AN IN-DEPTH BEHAVIORAL ANALYSIS OF FRAUDULENT NFT CREATORS}

\author{Trishie Sharma}
\email{tsharma21@iitk.ac.in}
\affiliation{%
  \institution{Indian Institute of Technology}
  \city{Kanpur}
  \state{Uttar Pradesh}
  \country{India}
  \postcode{208016}
}

\author{Rachit Agarwal}
\affiliation{%
  \institution{Merkle Science}
  \city{Bangalore}
  \country{India}}

\author{Sandeep Kumar Shukla}
\affiliation{%
  \institution{Indian Institute of Technology}
  \city{Kanpur}
  \state{Uttar Pradesh}
  \country{India}
}

\renewcommand{\shortauthors}{Sharma, et al.}

\begin{abstract}
  The explosive growth of non-fungible tokens (NFTs) on Web3 has created a new frontier for digital art and collectibles, but also an emerging space for fraudulent activities. This study provides an in-depth analysis of NFT rug pulls, which are fraudulent schemes aimed at stealing investors' funds. Using data from 758 rug pulls across 10 NFT marketplaces, we examine the structural and behavioral properties of these schemes, identify the characteristics and motivations of rug-pullers, and classify NFT projects into groups based on creators' association with their accounts. Our findings reveal that repeated rug pulls account for a significant proportion of the rise in NFT-related cryptocurrency crimes, with one NFT collection attempting 37 rug pulls within three months. Additionally, we identify the largest group of creators influencing the majority of rug pulls, and demonstrate the connection between rug-pullers of different NFT projects through the use of the same wallets to store and move money. Our study contributes to the understanding of NFT market risks and provides insights for designing preventative strategies to mitigate future losses. 
\end{abstract}

\ccsdesc[500]{Security and privacy~Economics of security and privacy}
\ccsdesc[300]{Security and privacy~Social aspects of security and privacy}
\ccsdesc[500]{Security and privacy~Privacy protections}
\ccsdesc[500]{Security and privacy~Pseudonymity, anonymity and untraceability}
\ccsdesc[300]{Security and privacy~Web application security}
\ccsdesc[300]{Information systems~Web applications}

\keywords{blockchain, cyber frauds, illicit activities, non-fungible tokens}


\maketitle

\section{Introduction}
Blockchain technology and Non-Fungible Tokens (NFTs) enable content creators (such as artists) to receive financial compensation for their work without relying on auction houses or galleries. An artist can now simply sell their work online in the form of NFT, aiming at increased reachability and acknowledgment. NFT marketplaces (NFTMs) are online platforms that facilitate the buying and selling of NFTs by charging sellers a fee for transferring non-fungible tokens (NFTs) from one party to another. NFTMs also provide additional tools for quickly creating NFTs. As the popularity and value of NFTs continue to grow, the number of frauds in the ecosystem will likely increase. The rise in frauds in the NFT ecosystem has become a significant concern for investors and stakeholders in the market~\cite{Chainalysis}. Fraudulent activities, such as rug pulls, hacking, phishing, and market manipulation cause significant financial losses and undermine the trust and stability of the NFT market. In 2021, victims lost more than USD 2.8 billion in cryptocurrencies to rug pulls. Rug pulls become the DeFi ecosystem's go-to fraud, generating 37\% of all bitcoin scam earnings in 2021 compared to just 1\% in 2020~\cite{Chainalysis}.

Suppose an NFT project advertises itself as an excellent investment scheme with the potential for huge returns and draws investors to support it. Creators typically include a roadmap before beginning an NFT project, describing their long-term objectives, often entailing the development of a future online game or earning money for charitable organizations~\cite{pixelmonCase,frostiesCase}. Then the NFT projects engage in auctions, pre-sales, mints, or airdrop campaigns to raise funds to complete the stages of their roadmap. However, many con artists use this fascination to lure investors into ``buying in'' to their new project and steal the received funds by shutting the project down immediately after receiving funds. Thus, these creators of an NFT project are said to have \textbf{\textit{Rug pulled}} when, after amassing enough funds, they dump the entire project and vanish with the money~\cite{RP1}. 
The creators deactivate all their social media presence, delete all the endeavor records~\cite{EllipticNFT}, and disconnect themselves from the Internet. In such cases, the investors cannot track the creators through social media accounts or other means of expressing their concerns. 

The transaction patterns of a rug-pulled NFT project are not different from those that are not rug-pulled. The usage of investors' money by NFT creators marks the difference between the active and the rug-pulled NFT project. Since it is challenging to differentiate rug pulls from other crypto-crimes, it sometimes accounts for a more significant proportion of losses~\cite{Elliptic}. However, given the transparent nature of NFT project contract addresses, it is relatively easy to monitor suspicious outflows. 
Projects have been called out online for inexplicable blockchain activity, such as using investor funds to purchase other NFTs or sending sizable portions of funds into \textit{exchanges, mixing services, swapping services, or illicit accounts}. Such activity indicates that investor funds are not being utilized in a manner that would realize the project’s roadmap\cite{EllipticNFT}. This highlights the need to thoroughly analyze transactions involved in rug-pulled projects to understand criminal intentions and behavior. 

Rug pulls have been studied previously, where researchers have examined the ERC-20\cite{erc20} tokens on decentralized exchanges to identify features for detecting rug pulls\cite{doNotRugOnMe}. Investigating the cryptocurrency trading on Uniswap\cite{uniswap} by collecting all the Uniswap V2 exchange transactions and analyzing them from different perspectives has enabled the identification and analysis of scam tokens\cite{TradeorTrick}. The cases studied on NFTs suggest introducing a register of utility NFTs for verifying the project's authenticity to lower the rate of rug pulls\cite{registryTool} or including scam prevention training and reporting any fraud to security agencies related to crypto-crimes\cite{CryptocurrencyFraud}. This research is limited to single case that must be revised for justifiable inferences, hence, no serious effort to analyze rug pulls on NFTs. It is challenging to discover the relevant transactions to categorize an NFT project as a rug pull\cite{Elliptic}. This study is a novel attempt to fill the gaps by providing essential insights from a dataset of 727 rug pulls on the OpenSea marketplace to aid in creating efficient rug pull detection systems.
We analyze the incidents of rug pulls and find the similarities in creator(s), shared accounts, and flow of funds. We categorize the 727 NFTs into 21 groups to identify correlations among creators and examine the dynamics of the largest group. We discover that the most common behavioral pattern in which creators collaborate to attempt rug pulls is the use of dedicated shared accounts. Our research gives insights into extracting features related to the timeline of an NFT project activity or NFT trading transactions for detecting a rug pull. In the following sections, we explore rug pulls and their key characteristics, methods used for data collection and analysis, the essential findings of this work, and summarized conclusions of this study.

\subsection{Rug pulls}
\begin{figure}[H]
    \centering
    \includegraphics[scale=0.29]{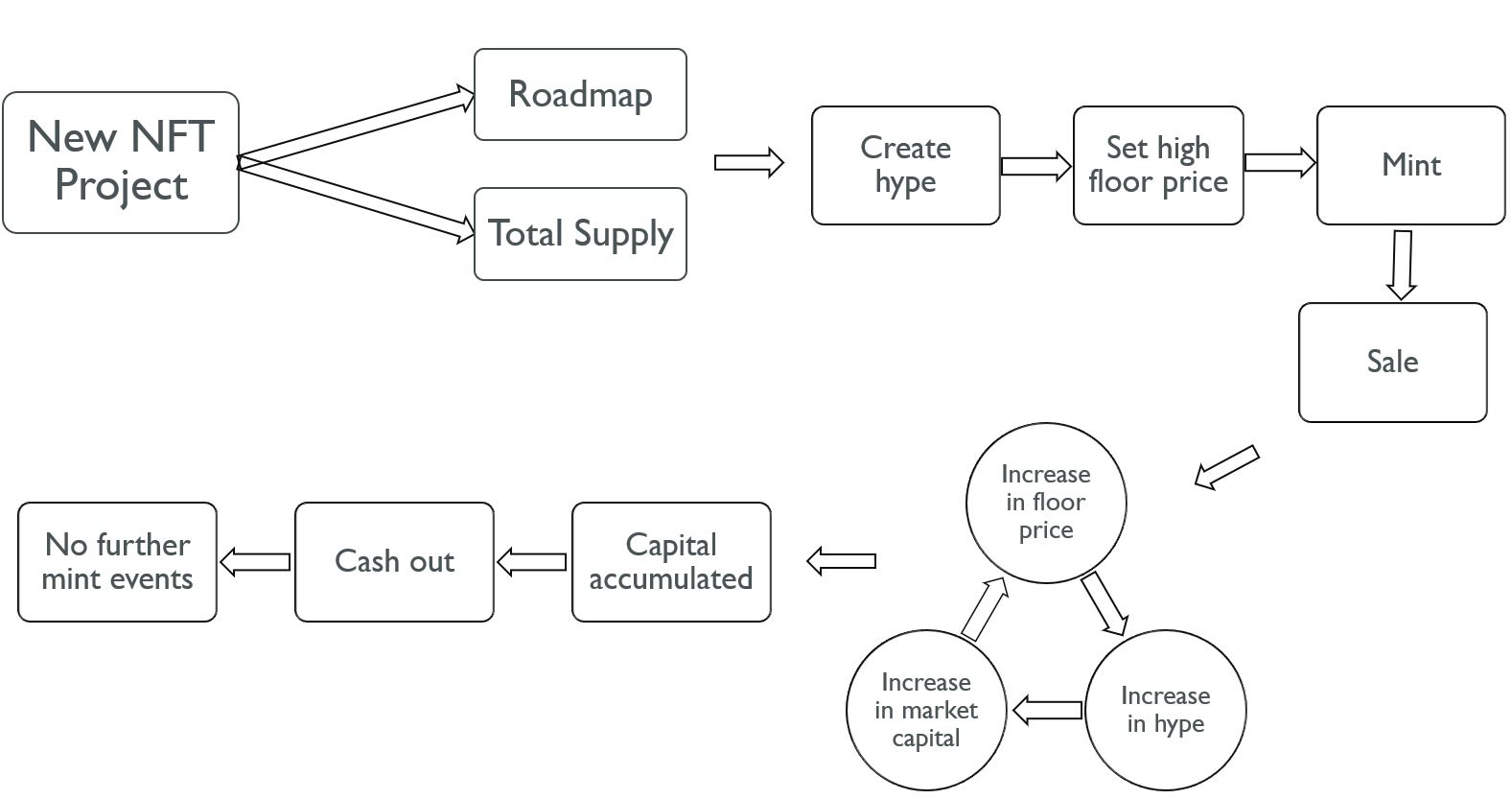}
    \caption{An example scenario of how a rug pull takes place}
    \Description{In a typical case of rug pull, an NFT project is created with its roadmap and total supply of tokens. Various tactics are used to make the project popular and a floor price is set. NFTs are minted and listed for sale which accumulates more capital with increase in popularity. Upon collecting sufficient funds, the creators cash out and abandon the project after taking away investor's funds.}
    \label{fig:1}
\end{figure}

Figure \ref{fig:1} depicts a typical rug pull scenario, beginning with creating a new NFT project that adheres to the ERC-721 smart contract standard\cite{erc721}. The ERC-721 standard introduces a non-fungible token standard where a token is unique and distinguishable from another token. These tokens are generated using a smart contract which is a unique address on the chain that provides methods describing an NFT project's permissible actions and deployed behavior. Specifically, every NFT holds a \textit{uint256} variable named \textit{tokenId}. Contract address and tokenId are a globally unique pair\cite{Wang2021NonFungibleT}. A smart contract is either generated by the project creators or by an NFT contracting firm\cite{nftContractingAgencies} that supplies the creators with a completed project. The project's creators publish a roadmap detailing the schedule of events and the \textit{total supply} that defines the total number of tokens that will be released into the market. Then, they establish a \textit{floor price}, the minimum amount an investor must pay to hold the NFT. The tokens are then produced, introduced to the market, and advertised for sale. Here, there are two elements of sales: the direct sale, which is the initial sale of an NFT, and the secondary sales, which include all the subsequent sales of that NFT. As sales increase, the market valuation of the NFT collection also increases. In a typical scenario of rug pull, after minting one or more batches of NFTs, the developers stop minting fresh tokens and transfer the proceeds to their wallets before ultimately abandoning the project.

The practice of trading NFTs involves promoting an NFT project on Twitter,  Discord and other forms of social media, other than simply listing NFT collections on NFT exchanges. Using social media to promote an NFT project is a tactic known as shilling, which is widely employed. \textit{Shilling} an NFT involves promoting it to persuade investors to purchase it, which is permissible regardless of ethics. Twitter is rife with shilling, where internet influencers, famous people, and celebrities promote NFT collections by posting videos supporting its founders and boasting about the project's strong return on investment potential. An NFT project's creators trade through legal and illegal means. In a rug pull, however, criminal activities prevail. 
One of the most prominent methods for promoting an NFT collection using misleading information is \textit{artificial inflation}, i.e., manipulating the numbers to indicate that a project is quite popular. For example, the creators of an NFT project create several accounts and then transact using those accounts to provide the appearance of a more significant number of investors. 

There is no definitive method by which creators execute a rug pull. 
To carry out such illegal acts while investors stay oblivious, creators mainly employ strategies that lead to four types of rug pulls \cite{rpf1}. Creators deactivate their Discord, Twitter accounts, and websites after withdrawing the funds, and the NFT owners are barred from asking questions on Discord. This strategy is termed as \textit{hard rug pull}. The method in which the creator depletes a project's liquidity to render it unsustainable is known as a \textit{soft rug pull}. It is a \textit{slow rug pull} when the components of a roadmap do not deliver as promised, yet the project remains active over time. When a project includes several direct wallets or transaction links to previously reported rug pulls, it is referred to as \textit{rug pull by extension}. In this case, the creators create many NFT projects (primarily in a short period) using similar code and cash out after amassing a substantial sum. Such an approach of attempting multiple rugs pulls by the same person or a group using similar strategies is referred to in our work as \textit{repeated rug pull}, discussed in Section~\ref{ssubsec:rrp}.


\section{Methods}

In 2020, tweets on the NFT application Valuables (developed by the startup CENT) documented the earliest rug pull incidents. The Valuables platform enables users to purchase a digital certificate for a tweet and sell it for Ethereum on the open market\cite{valuables}. We do not consider rug pulls related to tweets in this study. We collect details of NFT projects that are rug pulled on 10 NFT platforms. The term ``platforms" refers to NFT marketplaces and the NFT contracting agencies (that create NFT projects for content creators). The collected details include the project name, NFT platform, month and year of rug pull, creator, and contract addresses. Table \ref{table:1} provides a breakdown of rug pulls by platform and year between the collection period. The table depicts seven distinct NFT platforms, with SolSea representing all Solana blockchain-based marketplaces, including MagicEden\footnote{https://magiceden.io/}, Solanart\footnote{https://solanart.io/}, and DigitalEyes\footnote{https://www.digitaleyes.market/}.

\begin{table}[ht]
\begin{center}
\caption{Rug pulls on NFT platforms (June 2021 to November 2022)} 
\begin{tabular}{|l|c|c|c|} 
\toprule
\textbf{Platforms} & \textbf{2021} & \textbf{2022} & \textbf{Total} \\
\midrule
\textbf{AstraLabs} & 0 & 1 & 1 \\
\textbf{HashAxis} & 0 & 2 & 2 \\ 
\textbf{LooksRare} & 1 & 1 & 2 \\ 
\textbf{OpenSea} & 233 & 494 & 727 \\ 
\textbf{Rarible} & 0 & 2 & 2 \\ 
\textbf{Solana} & 9 & 14 & 23 \\
\textbf{WAX} & 1 & 0 & 1 \\
\midrule
\textbf{Total Cases} & \textbf{246} & \textbf{517} & \textbf{758} \\ 
\bottomrule
\end{tabular}
\label{table:1} 
\end{center}
\end{table}
 
AstraLabs\footnote{https://astralabs.io/} is an NFT contracting agency that assists content providers with developing and selling their NFT collections. AstraLabs, LooksRare\footnote{https://looksrare.org/} and Rarible\footnote{https://rarible.com/} interface with Ethereum, whereas HashAxis\footnote{https://hashaxis.com/} and WAX\footnote{https://wax.atomichub.io/market} support Hedera Hashgraph\footnote{https://hedera.com/} and WAX\footnote{https://wax.io/} blockchains, respectively. OpenSea\footnote{https://opensea.io/} supports the following blockchains: Ethereum, Klaytn, Polygon, and Solana\cite{openseaPlatform}. 
Appendix Section~\ref{app:A} shows a timeline graph of the monthly number of rug pulls on all NFT platforms.

OpenSea, the most active NFT marketplace with over 1 million active users\cite{OpenseaStats}, reported 727 rug pulls. Due to the highest number of rug pulls and a 36\% rise in rug pulls between 2021 and 2022, our research is based on a novel dataset of transactions from NFTs listed on OpenSea Marketplace. We focus on NFTs managed by token contracts deployed on the Ethereum blockchain. All Ethereum blockchain transactions are public, and their records are accessible via many blockchain explorers, such as Etherscan\footnote{https://etherscan.io} or third-party APIs, such as Ethereum-ETL\footnote{https://ethereum-etl.readthedocs.io} and Infura\footnote{https://www.infura.io/product/ethereum}. Based on the NFT projects that are rug pulled, we compile a dataset including the details of each such instance between June 2021 and December 2022 from Etherscan. 
We collect details about the associated NFT projects. We include the NFT platform used for minting tokens, the NFT project's creation timestamp, the month and year the project was terminated, the cashed-out value (in \textit{ethers}), and the contract and creator addresses. Our dataset contains all smart contracts generated by addresses involved in rug pulls. After a rug pull, specific NFT projects are managed by the community or a new management team. In such situations, we collect only the transactions until a withdrawal turns the account balance to zero or a minimum value. There are 68,618 transactions involving 1,052 ERC-721 tokens in the dataset. Out of 1,052 such smart contracts, 727 ERC-721 or NFT projects are detected as rug pull. The transactions of the rug-pulled NFTs vary between 2018 to January 2023. 

To compile the dataset, we perform web scraping and manual analysis of rug pull incidents collected from various crypto-related websites, news articles, and Twitter accounts, including ZachXBT\footnote{https://twitter.com/zachxbt}, Rug Pull Finder\footnote{https://twitter.com/rugpullfinder}, NFT Ethics\footnote{https://twitter.com/NFTethics}, CryptoShields.eth\footnote{https://twitter.com/cryptoShields}. Note that the users of these Twitter accounts usually find and report rug pulls and provide in-depth analyses to back up their assertions and validation. We study their data and confirm its accuracy by evaluating the transaction behavior of those projects. Via other public sources (such as Etherscan) that also provide attribution to such projects, there is no way to identify rug-pulled addresses. However, Etherscan labels some addresses implicated in frauds as `Fake Phishing' and `Spam.' The lack of built-in labeling necessitated a hybrid approach incorporating web scraping and manual evaluation.

For more in-depth analysis, we divide and label addresses into six entities according to their transaction behavior: (i) creators (CR), (ii) Externally Owned Accounts (EO - other accounts held by creators), (iii) illicit accounts (IL), (iv) smart contracts (SC), (v) temporary accounts (TE), and (vi) Virtual Asset Service Providers (VA).
Here, creators are the addresses that generate a smart contract for creating NFTs. We use the term \textit{wallet} to refer to other Externally Owned Accounts (EOAs) or addresses held by creators. 
We identify certain accounts as \textit{temporary accounts} and divide them into two categories based on their behavior. One kind corresponds to accounts with identical transactions of receiving funds from one or more accounts and immediately sending them to exchanges or sending funds in batches. These accounts typically have less than five transactions. The second type of temporary accounts are addresses that receive funds from certain specific addresses and use those funds to interact with other smart contracts for minting, purchasing, and trading activities.
Due to its involvement as a funds transmitter, Anti-Money Laundering/Combating the Financing of Terrorism (AML/CFT) and other requirements may apply when a digital asset firm engages in certain financial transactions involving virtual assets. These entities represent digital assets and are known as \textit{Virtual Asset Service Providers (VASPs)}\cite{VASPs}. We use VASPs to represent exchanges, mixers, swap services, and other services that fall under the concept of VASPs. Other services include wallet security providers, multi-sender applications, airdrop helpers, liquidity providers, crypto donation addresses, and marketplace royalty services. 
We also consider royalties from OpenSea or Rarible as VASPs. 
\textit{Royalties} are received from OpenSea as earnings fees for each sale of the NFT, allowing creators to be rewarded fairly for their digital work.

A crypto mixer is a service that blends the cryptocurrencies of many users to obfuscate the origins and owners of the funds\cite{mixers}. Swapping is exchanging crypto assets for their equivalent value in another coin or token\cite{swap}. In this work, we came across only one and the most used Ethereum mixer service, Tornado Cash\cite{tornadocash}. Note that Tornado Cash is sanctioned by U. S. Treasury\cite{tornadocash} and thus an illicit entity. Swap services have also become a significant cash-out and money-laundering tool for cybercriminals. Illicit variants of swap services promote them based on how “clean” the received funds will be and charge extra to swap crypto from illicit sources\cite{mixerswapillicit}. 

Besides mixing and swapping services that are considered illicit, other illicit activities exist and have associated addresses. For this work, we define illicit addresses as those involved in phishing or spamming. These addresses are identified by labels `Fake Phishing' and `Spam' on public explorers such as Etherscan. Further, other websites such as \cite{scammerAddresses} also help us enrich our illicit address data. We know that other sources (such as CryptoscamDB) provide information about an address being illicit or not and other services (such as Darknet services and gambling services). However, we do not use them due to resource constraints, and we only want to focus on one other category besides mixing and swapping. Further, note that to observe the transaction patterns and flow of funds, we consider `mixers or swap services' as VASPs and not `illicit addresses.' 


\section{Results}

We identify the most frequent behavioral patterns across all NFT rug pull projects and present our inferences on the behavioral patterns of rug pull projects using a series of associated transactions. Let `A' be the funding source address that transfers the funds to another address, `B.' Using those funds, the owner of address `B' creates an NFT project. Further, address `B' accumulates funds from the NFT project, which are then used for trading at VASPs. Note that exchanging money at VASPs by an NFT creator does not necessarily signal a rug pull; nevertheless, a significant decline in the number of transactions after a withdrawal implies inactivity and, ultimately, the abandonment of the NFT project. 

\subsection{Behavioral patterns}\label{ssubsec:behav}

For behavioral patterns, we focus only on those methods listed in the smart contract that are associated with money transfer, i.e., \textit{deposit()}, \textit{transfer()}, \textit{release()} and \textit{withdraw()} methods in different forms. By different forms, we mean different method names used by different creators in their smart contracts. These methods result in transferring funds from one address to the other. The \textit{deposit()} and \textit{transfer()} methods correspond to receiving and sending money, respectively, and the \textit{withdraw()} and \textit{release()} methods are called to transfer the funds accumulated by the NFT into its creator's account or any other account used by the creator. Apart from the methods related to the transfer of funds, we also show the creation of NFT, which refers to the methods using which a creator creates an ERC-721 smart contract. 

\begin{figure}[H]
    \centering
    \includegraphics[scale=0.58]{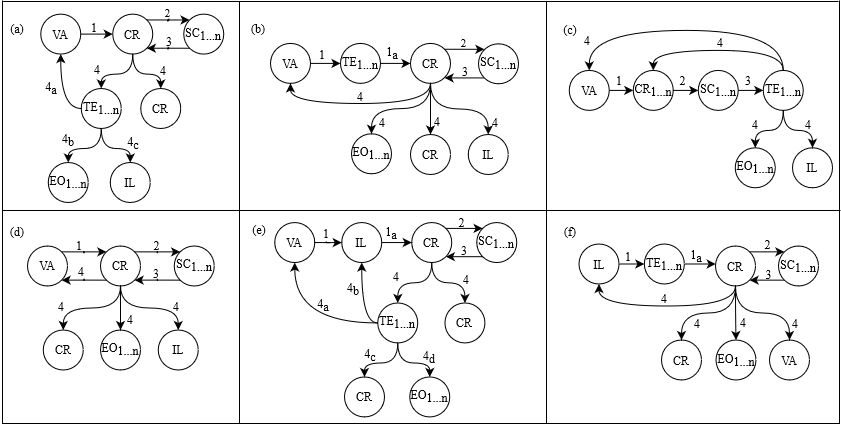}
    \caption{Prominent behavioral patterns observed from the flow of transactions in rug pulled projects. Here, 1, 2, 3 and 4 represent the four types of interactions, depicted as edges. CR (creator), EO (creator's wallet), IL (illicit address), SC (smart contract), TE (temporary account) and VA (VASPs) are the six entities. The interaction $1_a$ denotes transfer of funds from VASPs to a creator through an intermediary. The notations $4_a$, $4_b$, $4_c$ and $4_d$ indicate that their parent node serves as a mediator for the transfer of funds to other entities.}
    \label{fig:2}
    \Description{Six most common patterns identified from creator's transactions involving flow of funds.}
\end{figure}

From our dataset, we create a directed graph depicting the interactions of all rug-pulled project creators (cf. Appendix Section~\ref{app:b}).
From this complete graph, we extract and show creators' most prominent behavioral patterns in figure \ref{fig:2}. Appendix Section~\ref{app:c}
details the patterns associated with each of the six entities. Figure \ref{fig:2} shows the network where edges represent transactions using the above-mentioned methods. Since this work highlights the transaction patterns of rug-pulled projects, we emphasize only the transactions related to the flow of funds. Hence, we do not consider other methods such as minting, sale, purchase, and others related to token contract operations. We identify four types of interactions in these patterns based on our comprehensive analysis of transactions. 

\subsubsection{Interaction 1: Initial funds} symbolizes the very first source of VASP funding. Finding royalties as the first income source for a creator suggests a connection to the creator's earlier project(s), for which the creator got paid. Funds obtained through an exchange, mixing, or swapping services also indicate connections to previous projects. Such behavior suggests that the funds have been acquired by trading past NFT projects' earnings. The only way to evaluate this option is to examine the exchange's transactions, which is beyond the scope of this study. Figures \ref{fig:2}(b) and \ref{fig:2}(f) depict the behavior where the funds flow through one or more of the creator's temporary accounts, $TE_{1...n}$. 
Here, the temporary accounts act as an intermediary between VASPs and creator (cf. Figure \ref{fig:2}(b)) and illicit address and creator (cf. Figure \ref{fig:2}(f)), hence, considered as a subset of interaction 1 and denoted as $1_a$.  
Similarly, figure \ref{fig:2}(e) depicts the flow of funds with an illicit account being the intermediary. Figure \ref{fig:2}(c) depicts another approach to funding a creator via one or more creators, represented by $CR_{1...n}$. Figure \ref{fig:4} shows this relationship between different creators. 

\subsubsection{Interaction 2: NFT creation} signifies the issuance of a token contract for the NFT utilizing the funds acquired in Interaction 1. A token contract is a smart contract created by the marketplace or the NFT creator, including the methods for managing NFTs\cite{understandingSecIssues}. It is worth mentioning that a single creator can generate several NFTs, denoted by $SC_{1...n}$. 

\subsubsection{Interaction 3: Withdrawal of funds} exemplifies that after NFT creation (Interaction 2), it is introduced to the market to generate capital from investors or purchasers. After a period, the acquired funds are withdrawn by the creator.

\subsubsection{Interaction 4: Trading of funds} indicates the transfer of withdrawn funds to VASPs, illicit accounts, one or more wallets of the creator, indicated by $EO_{1...n}$, or other creators. In the case of rug-pulled NFTs, these EOAs are used to collect funds and do not have any outgoing transactions. The creator is presumed to employ these addresses for activities such as launching a new NFT project and using them as a temporary account for transferring funds to VASPs. Figures \ref{fig:2}(a) and \ref{fig:2}(e) demonstrate the behavior in which temporary accounts, $TE_{1...n}$, are utilized by their creators to transfer money. The formation of trails from multiple temporary accounts of a creator is a pattern frequently found in the transactions.

Using the transaction graph, we capture patterns in the flow of funds to demonstrate where and how the flow of funds of a rug-pulled project end. These patterns also serve as a means of establishing connections between various creators of NFT projects involved in a rug pull. One of the most prevalent instances of this relationship is when one creator finances the other. Over time, after accumulating capital, the other creator sends back the same and sometimes more funds. Figure \ref{fig:4} demonstrates such a sequence of transfers between multiple creators, whose interactions result in diverse forms of structures, as described in Section \ref{ssubsec:creators}. These structures further illuminate repeated rug pulls, explained in Section \ref{ssubsec:rrp}. The incoming funds from illicit accounts help identify the creator addresses of those responsible for the rug pulls (cf. Section \ref{ssubsec:rpgroups}). 
We discover many creator addresses that moved all or a portion of funds to illicit addresses, transferring those funds to mixing services over time.
We also explore the flow of funds between creators and other entities (exchanges or services, mixers or swap services, illicit accounts, and temporary accounts) in Appendix Section \ref{app:d}.

\begin{figure}[H]
    \centering
    \includegraphics[scale=0.185]{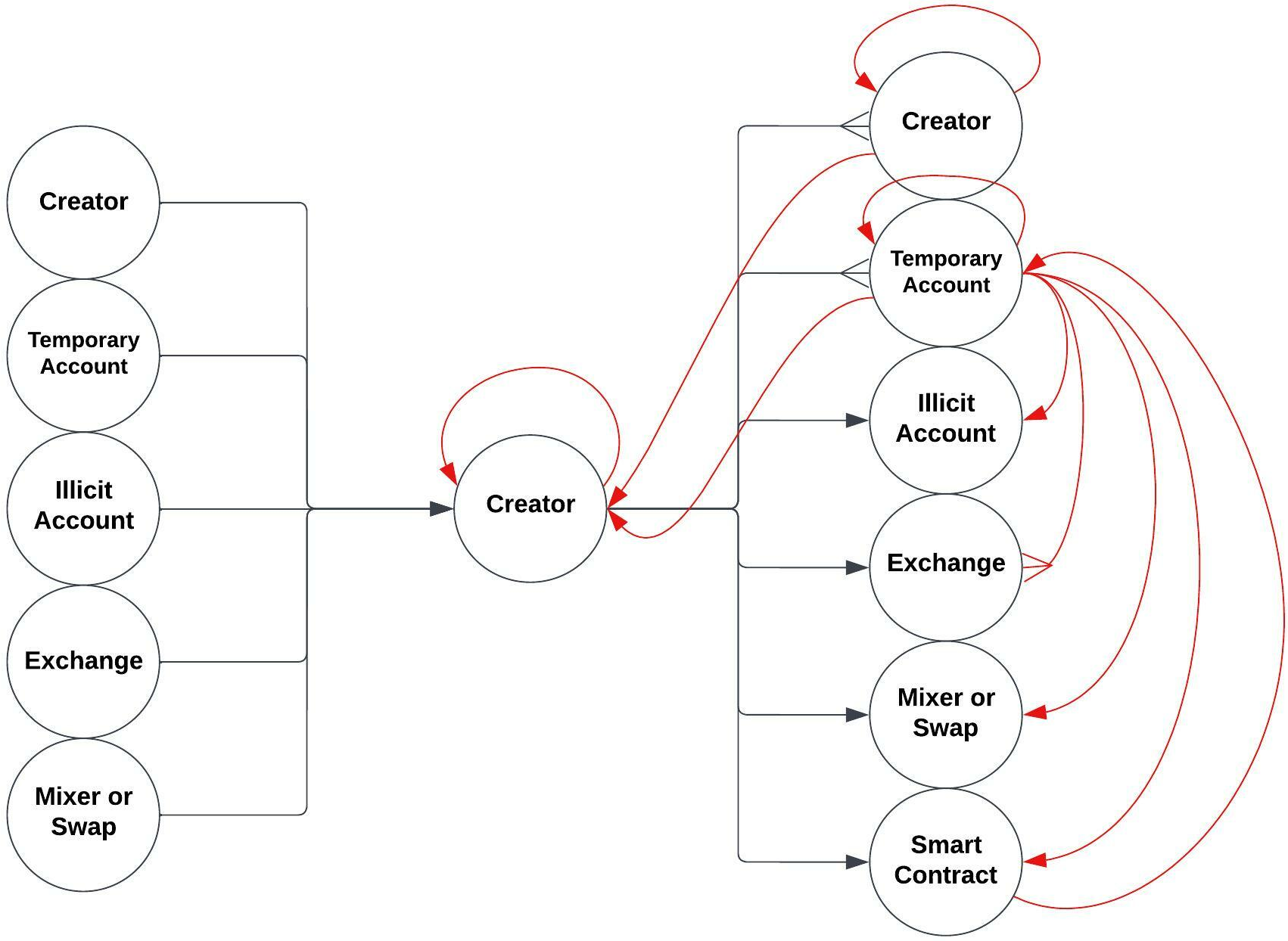}
    \caption{A holistic view of the transaction patterns of a rug pulled project creator. Black arrows represent the first transaction. Red arrows indicate all the subsequent transactions. The fork arrows denote one to many relationships.}
    \Description{Creators receive funds from creators, temporary accounts, illicit accounts, exchanges, mixers or swaps and transfer funds to all the six entities.}
    \label{fig:12}
\end{figure}

Exchanges, mixers, and swap services fall under the category of VASPs, as mentioned earlier; however, to understand and illustrate the interactions of `exchanges' and `mixers or swap services' separately, we consider them as two distinct entities. We also consider creators and their corresponding wallets (EOAs) as a single entity. Following these considerations, figure \ref{fig:12} depicts the holistic view of transaction behavior based on all the observable patterns stated previously. 

Figure \ref{fig:12} shows the transaction flow patterns, beginning with the initial transaction of the creator, shown by black arrows, and ending with the last transaction recorded in our dataset. A red arrow depicts each successive transaction. 
We use crow's foot, or fork notation \cite{crowsFootNotation} to represent one-to-many relationships, such as transactions beginning with a single creator and extending to several creators or temporary accounts. Similarly, we represent the transactions from a temporary account to several exchanges using forked notation.

\subsection{Understanding the network of creators}\label{ssubsec:creators} 

\begin{figure}[H]
    \centering
    \includegraphics[scale=0.135]{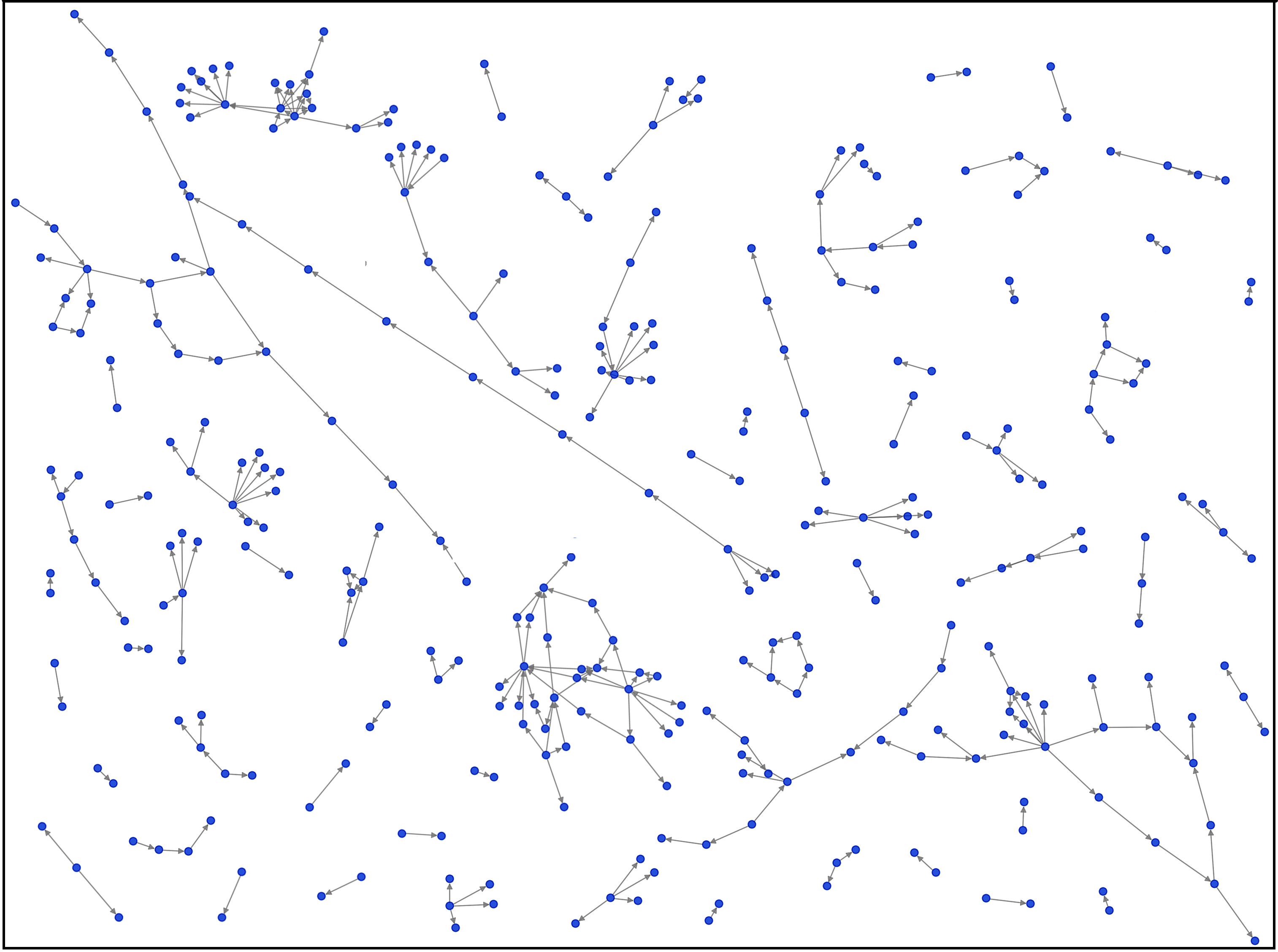}
    \caption{Direct interactions between creators of all rug pulled NFT projects causing formation of various clusters with different structures.}         
    \label{fig:4}
    \Description{Different chain-structures formed between creator addresses.}
\end{figure}

Different structures in figure \ref{fig:4} represent the movement of funds between the creator addresses. In most instances, multiple creators are behind one or more NFT projects, indicating repeated rug pulls (cf. Section~\ref{ssubsec:rrp}). These creators assist each other's NFT with inputs, such as finances or purchases, to demonstrate increased purchases, adding to the project's popularity and luring investors. The recipient creator then utilizes the funds to cover any transaction or gas fee (fee paid to the miners for inclusion of the transaction in the block) expenses incurred in the transaction related to the creation of the smart contract or any other transaction. Later, creators use these funds for trading or creating further tokens. Transferring funds from one rug pull creator's address to another resembles a chain structure. A star structure implies that a single creator has created numerous tokens. We discover that rug pull projects last a week or two before cashing out and returning with a new token (cf. figure \ref{fig:5}). 

\begin{figure}[H]
    \centering
    \includegraphics[scale=0.1]{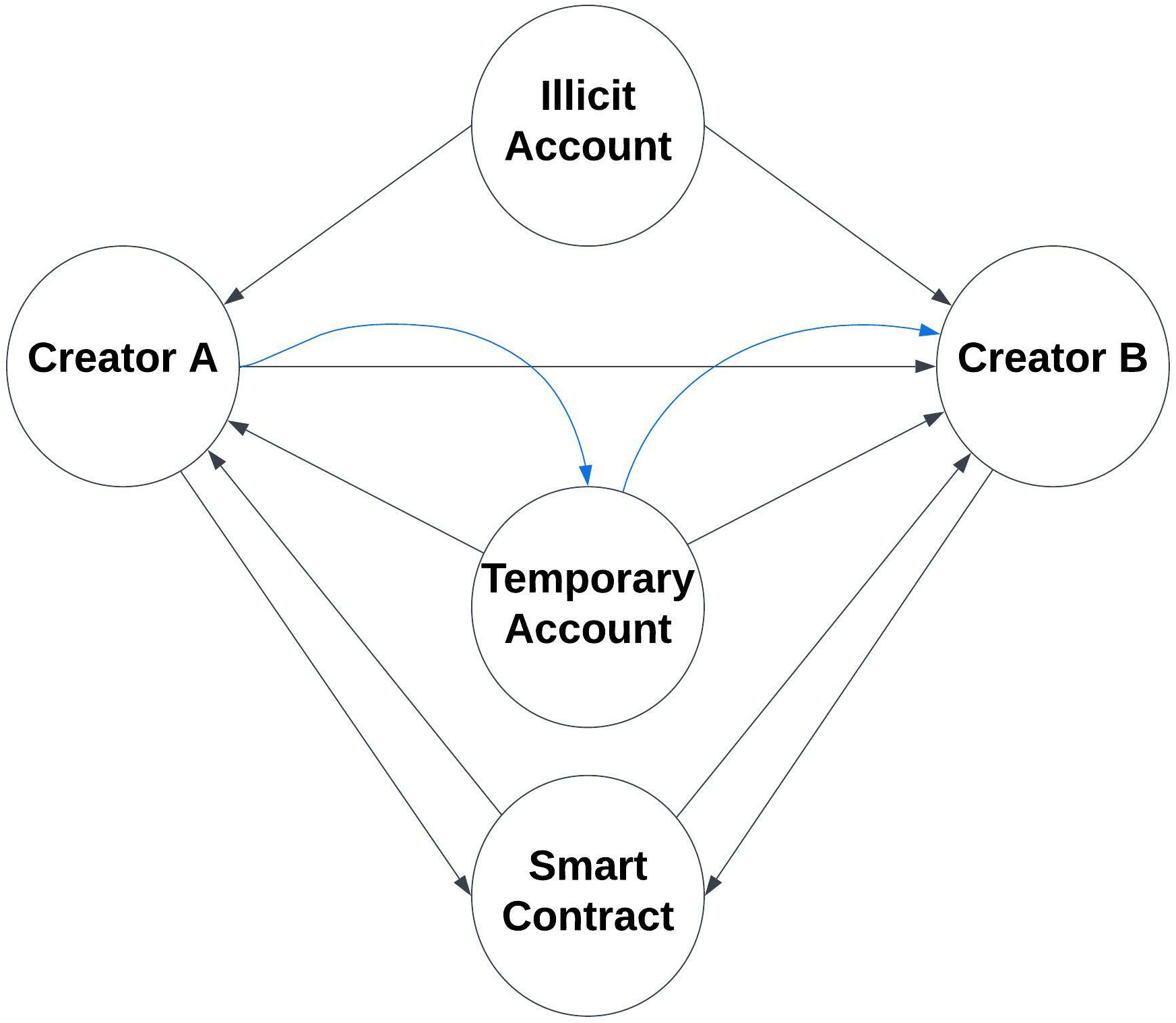}
    \caption{Means of connection between any two creators. Black arrows represent direct link. Blue arrows indicate usage of common temporary account of two creators.}
    \label{fig:11}
    \Description{Any two creators are connected through an illicit account, one or a trail of temporary accounts or a smart contract.}
\end{figure}

To understand the collaboration of creators to attempt rug pulls, we evaluate the interconnections between them. Figure \ref{fig:11} illustrates the interrelationship of any two creators. A connection between two creators is either through (i) one or more temporary accounts, (ii) an illicit account for initial funds, or (iii) a smart contract with similar interaction occurrences. However, the most common case involves creators operating from the same temporary account.

\subsection{Repeated Rug pulls}\label{ssubsec:rrp}

A small alliance of influential individuals engage in repeated malicious acts, simply interested in receiving all of the funds after minting \cite{nftEthics}. They then proceed to the next project with a similar setup and same approach. We note that 63\% of the creator addresses are unique, indicating that multiple NFTs are created using a single address. If any NFTs created by the same address are involved in a rug pull, other NFTs are also created with the same intention. 

\begin{figure}[H]
    \centering
    \includegraphics[scale=0.35]{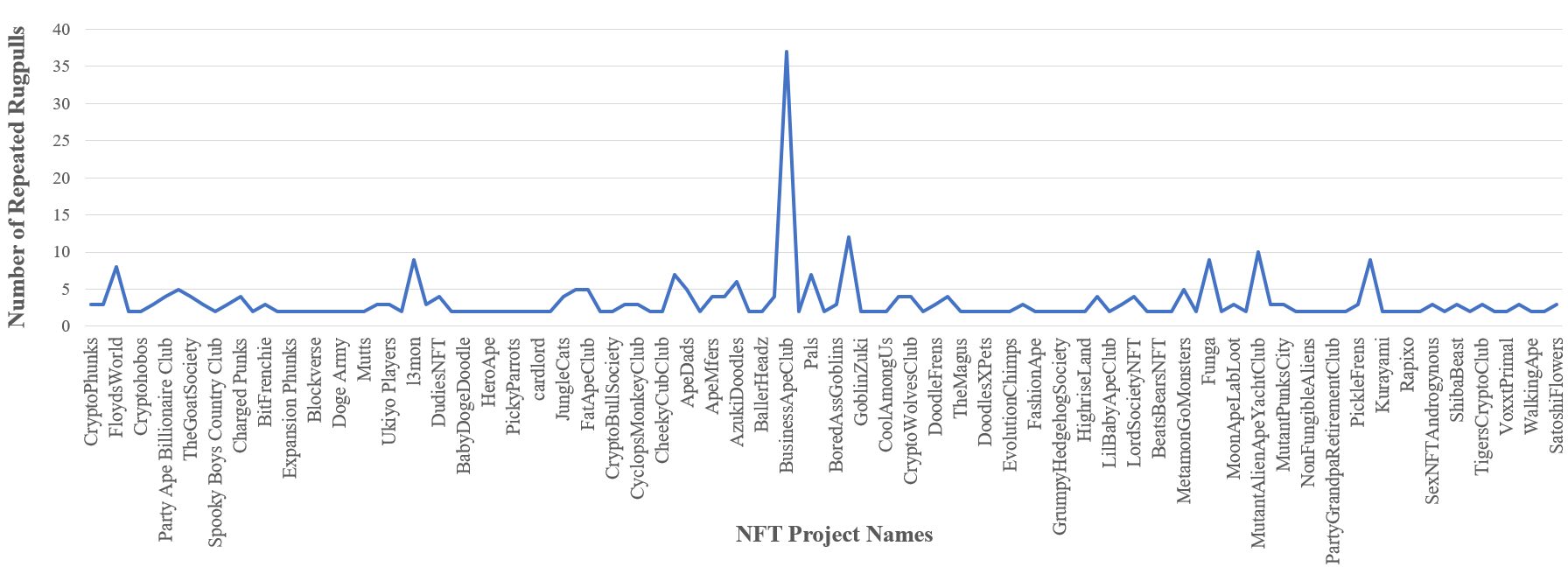}
    \caption{Number of times a creator created a project which ended up as a rug pull.}
    \label{fig:6}
    \Description{Line chart of all creators who attempted multiple rug pulls.}
\end{figure}

Figure \ref{fig:6} shows the NFT projects involved in repeated rug pulls. 42\% of projects on the OpenSea marketplace are identified as repeated rug pulls between June 2021 and December 2022, according to our data. In these instances, Business Ape, the creator of an NFT project, developed 38 smart contracts, of which 37 are NFT collections. These NFTs are created between December 2021 and February 2022, the maximum number of times a single creator has created NFTs to do repeated rug pulls. In another scenario, a team of scammers used similar techniques to take the investor's money before getting noticed. We found 20 such groups (cf. Section  \ref{ssubsec:rpgroups}).

\subsection{Dynamics of Rug pull Mafia}\label{ssubsec:rpgroups}

Our dataset contains 68,168 transactions among 458 unique creator addresses. By studying these transactions, we identify creators who created (i) multiple NFT projects from one account, (ii) multiple accounts, and subsequently multiple NFT projects using those accounts. Here, (i) and (ii) represent one-to-many and many-to-many relationships, respectively. Thus, we primarily focus on the direct interactions of the creators.

\begin{figure}[H]
    \centering
    \includegraphics[scale=0.45]{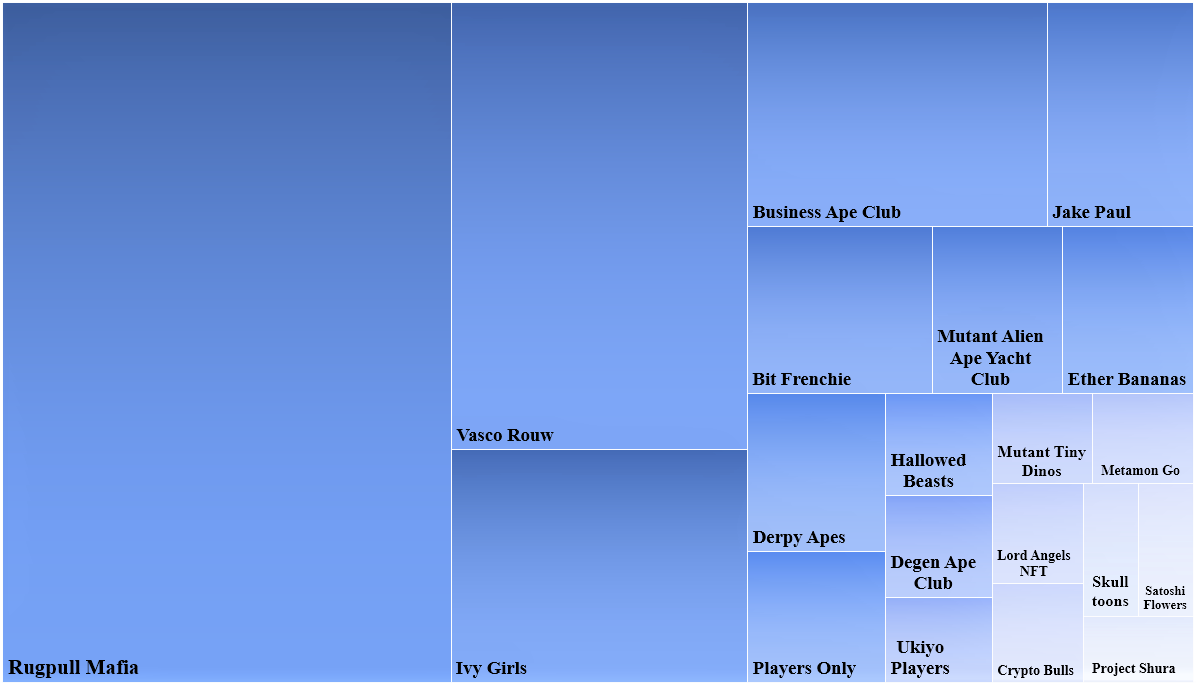}
    \caption{Proportion of 20 creator groups of creators who created multiple NFT projects. Some groups are named after the people linked to those projects.}
    \label{fig:9}
    \Description{Treemap chart of the categorized rug pull groups with their labels.}
\end{figure}

We identify 20 groups that exhibit this pattern. For the other remaining identified groups, we mark them all under the `Others' category. Thus have 21 groups in total. Figure \ref{fig:9} represents the relative proportions of each rug pull group based on the number of NFT projects in it. The `Others' group is not shown in figure \ref{fig:9} as the NFT projects are not distinctly identified.
We focus on the dynamics of the `Rug pull Mafia'\cite{rpmafia} group with the highest proportion (34\%) of all rug pull cases. This group has created 168 NFT projects as per our dataset. We study their transaction flow and find that all projects are interconnected in multiple ways, as discussed below.

\textbf{Source of funds from illicit accounts:} We found 25 illicit accounts in Rug pull Mafia group transactions as the initial funding source. Forty-five addresses received funds from the illicit accounts. The receiving addresses include 27 creator addresses, 16 temporary accounts, and two wallet addresses. Two more addresses are identified as the initial funding source for the 52 addresses, including 13 NFT project creators and six temporary accounts. The funded NFT projects further funded five more projects. The transactions associated with initial funding from illicit addresses occurred between December 2021 and May 2022. The same illicit and temporary accounts as the funding sources demonstrate a tight linkage across all NFT projects in context. It also shows that any new ERC-721 token generated by any of these addresses is likely to be a rug pull.

\textbf{Role of temporary accounts:} For creating NFT projects, funds are sourced directly from an illicit account, a VASP (including exchanges, mixers, and swap services), or another wallet related to the addresses of the projects. We discover 583 temporary accounts corresponding to addresses created deliberately to move cash from the creators to the exchanges. The average number of transactions in these accounts is less than 10. Some accounts have only three transactions: receiving funds from a source, minting an NFT, and transferring the NFTs to VASPs for further trading. These accounts also transfer funds to other temporary accounts and receive funds from creators. Intriguingly, these patterns occur within a month's time range and with the same creators. Moreover, as the number of project creations increased from September 2021 to February 2022, the number of transactions in such projects decreased to fewer than 30 and cashed out in less than 24 hours to 1 month (cf. Figure \ref{fig:5}). 

\begin{figure}[H]
    \centering
    \includegraphics[scale=0.22]{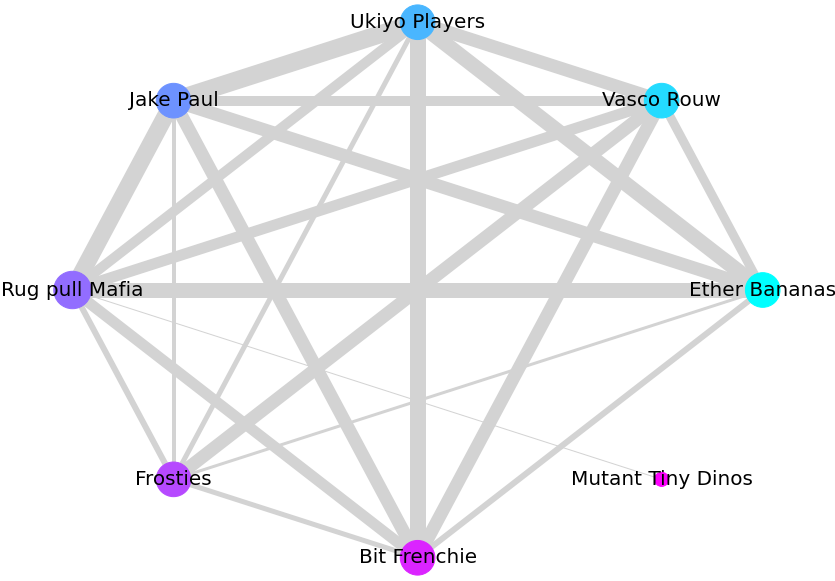}
    \caption{Connections between 8 rug pull groups identified from the common temporary accounts among them. Size of each edge represents the number of NFT projects associated with the common temporary accounts.}
    \label{fig:rugpullgroups}
    \Description{Connectivity graph of 8 rug pull groups. It shows relation between the groups. Different edge sizes indicate the common NFT projects between any two groups.}
\end{figure}

\textbf{Connections between creator groups:} To further investigate ties between the 20 identified groups, we analyze the indirect interactions between different rug pull groups. Transfer of funds to other creators or VASPs using the same temporary addresses demonstrates the association between different groups of creators. We look for those temporary accounts common in any two distinct groups. From the identified accounts, we look for the original NFT project of those temporary accounts. This gives the number of NFT projects in terms of connections between two rug pull groups. We discover that out of 20 rug pull groups, eight are linked with each other, creating 22 sets of associativity. Each set is: {groupA, groupB, number of common NFT projects}. We demonstrate the connections between the eight rug pull groups in figure \ref{fig:rugpullgroups}. The size of links between two nodes represents the number of common NFT projects that are identified. Association between creator groups suggests a strong collaboration in creating a notably large number of rug pulls.

\subsection{Timeline of creations and withdrawals in NFT rug pulls}
\begin{figure}[H]
    \centering
    \includegraphics[scale=0.65]{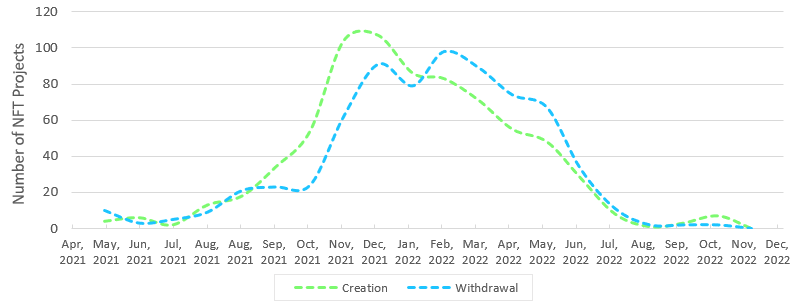}
    \caption{Comparison between the time of the creation of NFT projects according to the event of transferring investor's money to creator’s account upon attempting a rug pull.}
    \label{fig:creationcashout}
    \Description{Scatter chart with two lines. Green dashed line represents number of NFT projects created in a month. Blue dashed line represents the number of NFT projects that cashed out in a month. Time period is from April 2021 to December 2022.}
\end{figure}

\begin{figure}[H]
    \centering
    \includegraphics[width=\linewidth]{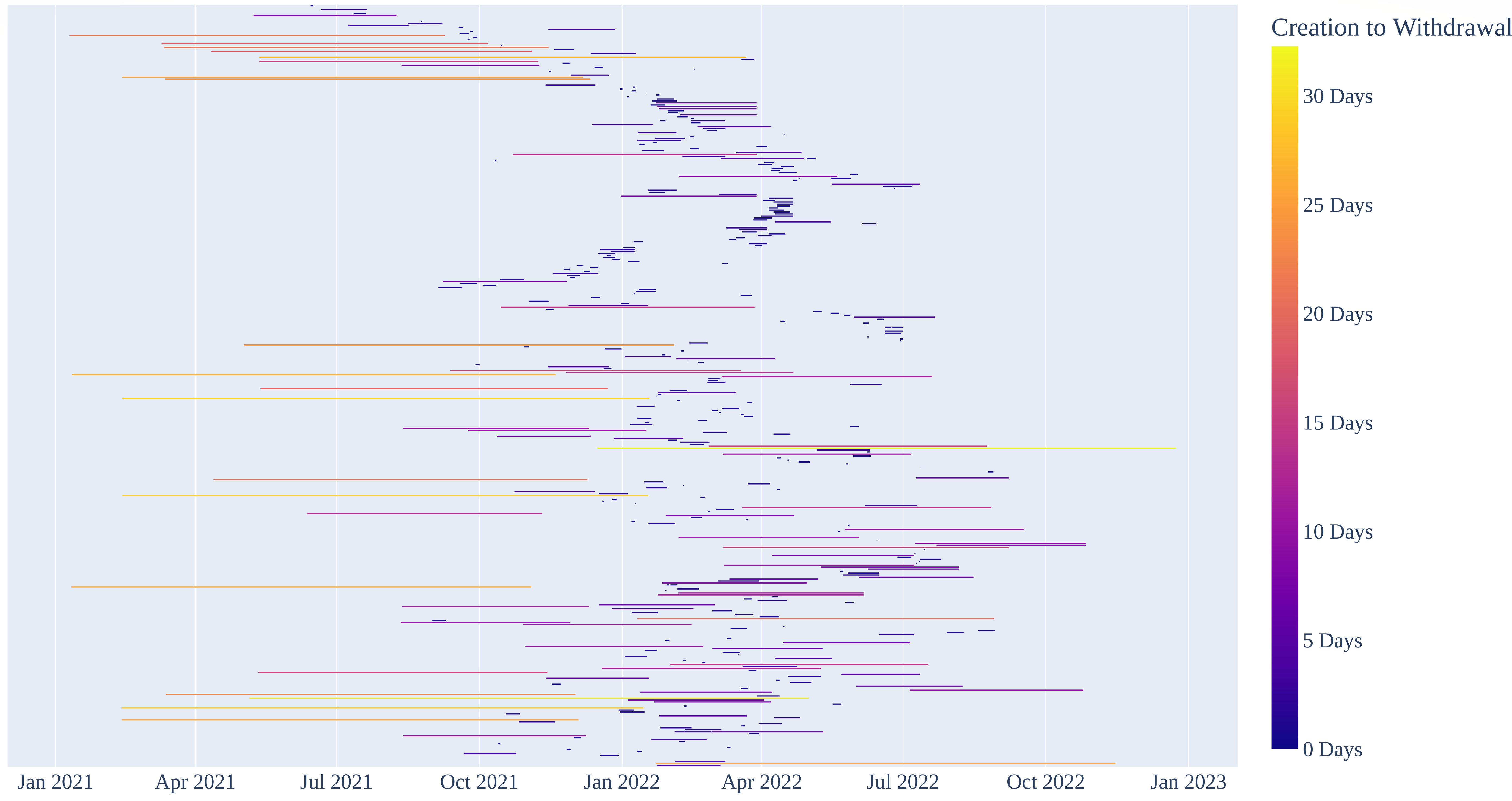}
    \caption{Timeline of the creation of NFT projects and withdrawal of funds. Each colored horizontal line depicts a rug pulled NFT project. The vertical colorbar represents the number of active days for a project before attempting a withdrawal, accounting for a rug pull.}
    \label{fig:5}
    \Description{Gantt chart of activity span of each NFT project. Activity span comprises creation and cashing out dates.}
\end{figure}

Figure~\ref{fig:creationcashout} shows the rise and decline in creating and cashing out of NFT projects with time. In the first quarter of 2022, more players preferred to sell quickly, for less price \cite{Nonfungible.com}. Similarly, an increasing number of NFT projects are abandoned quickly. We delve deeper into the creation of NFT projects and their fund withdrawals to understand the activity period of each rug-pulled NFT project (cf. Figure \ref{fig:5}). We find that 8\% of the projects rug-pulled within 24 hours of project creation and bringing the project to the NFT market. 
In figure \ref{fig:5}, lines represent the duration of NFT projects that are rug pulled between January 2021 and December 2022. Small blue lines and dots show this behavior. In 44\% cases (represented by blue and purple lines), the duration is less than a week, meaning the NFT projects are created and cashed out within a week.
Moreover, 37\% NFT projects lasted more than 15 days (represented by yellow and red lines). Thus, because 74\% of NFT projects lasted less than a month, it indicates that most of these projects are repeated rug pulls. Lawsuits against NFT creators and marketplaces, along with other factors, lead to a drop in the number of rug pulls after January 2021 (cf. Figure \ref{fig:creationcashout}). With the decreasing number of project creations, an increase in abandoning projects after cashing out funds is observed.

\subsection{Impact of involvement of law enforcement in curbing rug pulls}\label{ssubsec:creationwithdrawal} 
The fact that the creators of the NFT projects are not usually identified publicly and have a pseudonymous identity is a problem. Also, the number of token holders involved in a project is simple to manipulate which gives uninformed investors the idea that the project is either (i) genuine and in high demand or (ii) attracting investors and new users. This ease of manipulation and misdirection by creators indicates the lack of control and rules governing the workflow of NFTs and NFT markets. 

Beginning in November 2021, a surge in the creation of such projects is attributed to the lack of awareness among buyers regarding the identification of rug pulls, along with a solid desire to profit from investing in NFTs (cf. Figure \ref{fig:5}). The creators enticed customers by making tokens comparable to expensive ones but selling them at a reduced price or shilling, among other strategies. As people became more aware of scams through research and societal awareness, the victims began to seek assistance from various law enforcement agencies \cite{frostiesCase, floydkim, openseaLawsuits}, and the number of new rug pulls began to decrease. 







\section{Discussion}

Our research focuses on the incidents of rug pulls that occurred between June 2021 and December 2022 by analyzing a novel dataset of Ethereum transactions on the OpenSea marketplace. The collected rug pulls are similar in numerous ways and show similar dynamics, such as created by the same individual(s), usage of temporary accounts, and initial source of funds. In an attempt to correlate creators based on their transaction activity, 727 NFTs are categorized into 20 groups, including `Others.' 
There are 168 tokens linked with the `Rug pull Mafia' category, the largest group of creators responsible for multiple and repeated rug pulls. We focus our research on this group and study its dynamics to gain meaningful insights. Regarding the initial source of funds, the prevalence of temporary accounts to conceal creators' direct interactions with VASPs or illicit accounts, the activity period determined by their creation and cash-out dates, and the interconnectedness of several rug-pullers, we discover numerous parallels in the rug-pullers' behaviors. 

Since rug pullers use similar tactics, cloned or copied projects, we infer that it is simple for an average programmer to attempt a rug pull. We determine that cryptocurrency exchanges and OpenSea royalties are the principal funding sources. Using such funds to produce rug pull projects suggests that the preceding or prospective initiatives are also rug pulls with comparable transaction patterns. By gathering these observations and other commonalities, we seek to identify research opportunities for developing robust machine-learning models capable of detecting groups of creators attempting repeated rug pulls.

Our findings illuminate the transaction tendencies of the creators of NFT collections, which have been previously studied in the literature\cite{EllipticNFT, Nonfungible.com}. Intriguing future research directions include improving the analysis to determine whether regulated marketplaces affect rug pulls. Investigating the different types of transaction behavior aids in predicting a rug pull. Extending this research to include exchanges to determine the complete trail of fund transfers for each NFT collection is essential to detecting rug pulls. Also, including different types of illicit addresses (such as Darknet and gambling) to study their involvement in rug pulls is another direction to expand our research.

Besides rug pulls, analyzing transaction patterns will also help identify actors involved in money laundering and wash trading. Money laundering aims to legitimize money with illegitimate origins using various techniques\cite{moneylaundering}. Wash trading occurs when people trade assets between their accounts, usually to make interest in a given project appear higher than it really is\cite{washtrading}. Applying our behavioral analysis to identify malicious actors involved in money laundering and wash trading is another potential future work achievable by observing the presence of cycles\cite{NFTwashtrading} between creators and their related addresses in the rug pull transactions.
Our findings will pave the way for a better understanding of the self-organization aspects of emergent NFT marketplaces.

\begin{acks}
This work is partially funded by the National Blockchain Project (grant number NCSC/CS/2017518) at IIT Kanpur sponsored by the National Cyber Security Coordinator's office of the Government of India and partially by the C3i Center funding from the Science and Engineering Research Board of the Government of India (grant number SERB/CS/2016466).
\end{acks}

\bibliographystyle{ACM-Reference-Format}
\bibliography{biblio}

\appendix

\section{Timeline of Rug pulls}\label{app:A}
\begin{figure}[H]
    \centering
    \includegraphics[scale=0.45]{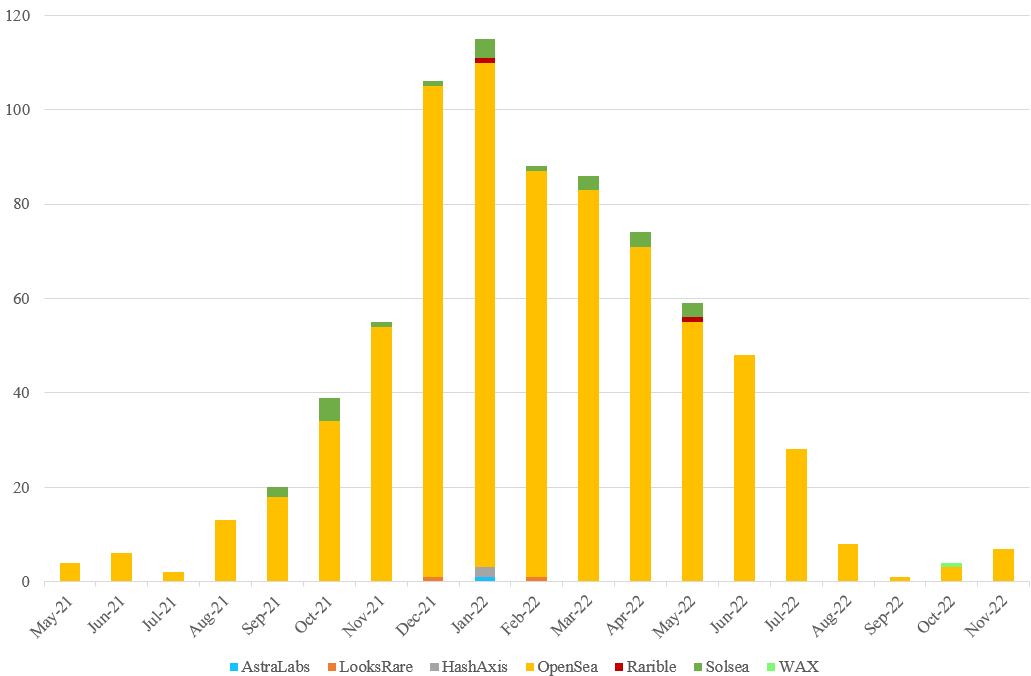}
    \caption{Timeline of Rug pulls on different NFT platforms in 2021-22}
    \label{fig:timeline}
    \Description{Column chart representing number of rug pulls that occurred on NFT platforms from May 2021 to November 2022.}
\end{figure}

From 2021 to 2022, the number of rug pulls increased by 35 percent. Figure \ref{fig:timeline} illustrates a significant rise in the number of rug pulls starting in December 2021. We observe this behavior to remain consistent for two months, followed by a decline beginning in April 2021. January 2022 saw the most rug pulls, with 93 percent of those occurring on the OpenSea market.

\section{Network of all transactions between different entities}\label{app:b}

\begin{figure}[H]
    \centering
    \includegraphics[scale=0.14]{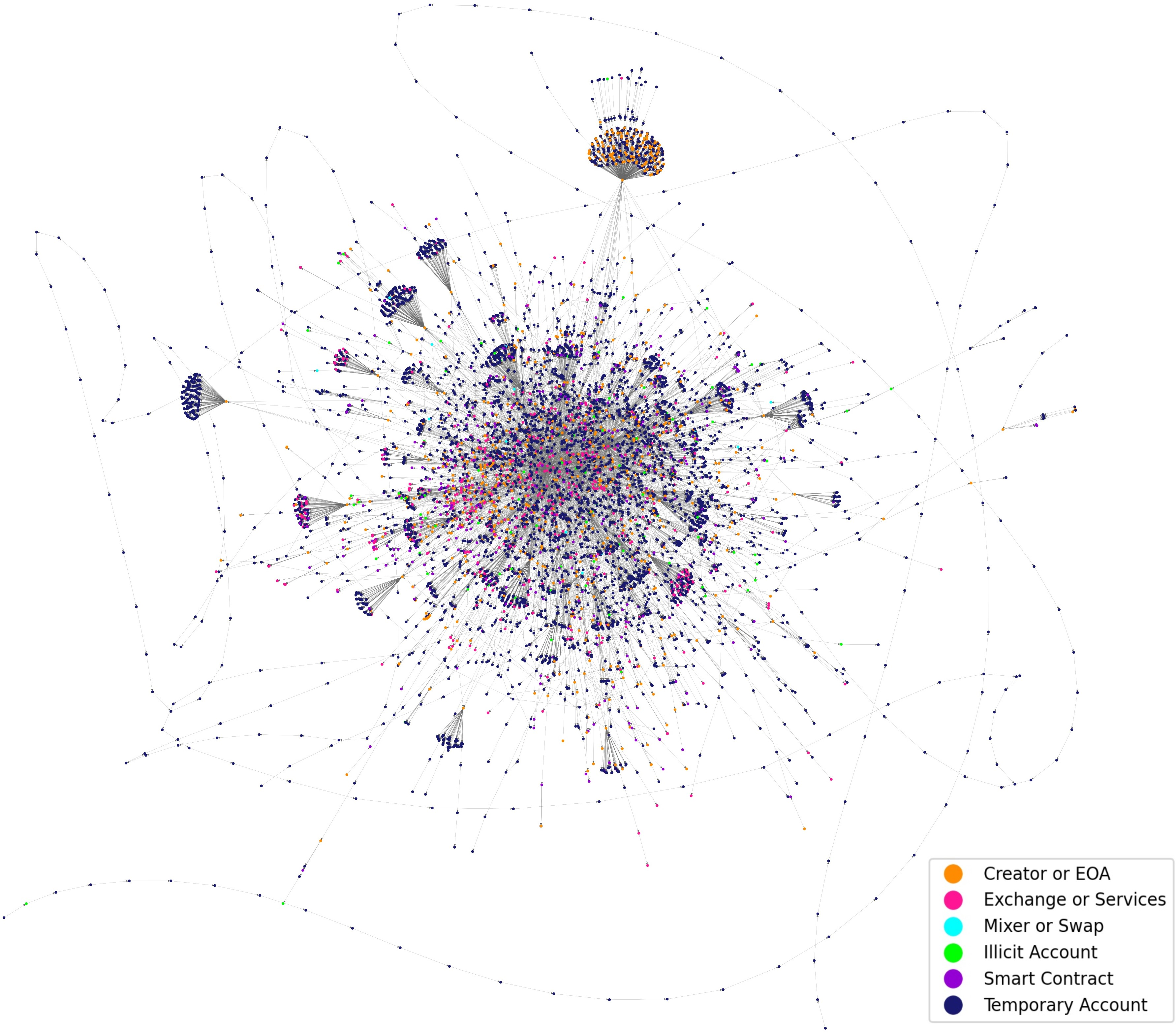}
    \caption{Flow of transactions among all the rug pull cases. Node colors are according to the type of entity: creator of an NFT or an externally owned account (orange), smart contract (purple), temporary account (dark blue). Dark pink nodes correspond to VASPs, whereas aqua nodes represent a mixer or swap service. Lime green nodes correspond to an illicit account.}
    \label{fig:allTxns}
    \Description{Graph created from the transactions recorded in our dataset. It shows direct transfers from one entity to another. The legend shows six entities: creator or EOA, exchanges or services, mixer or swap, illicit account, smart contract and temporary account.}
\end{figure}

To analyze the transaction patterns of creators, we create a network of creator's interactions with other five entities. Here, we consider creators and their other wallets (EOAs) that they own as a single entity and the other entities include exchanges or services, illicit accounts, mixers or swap services, smart contracts and temporary accounts. 
Figure \ref{fig:allTxns} displays the network of all interactions among the six entities. 
The network nodes are the six different entities considered in this work, while the edges represent the funds transfer. These transfers also include transfer of royalties from OpenSea.
The network displays usage of temporary accounts for transferring funds to exchanges, mixers, or illicit accounts via chain-like structures.
Smart contracts indicate a creator's call to the deposit, transfer, release and withdraw methods of NFT's smart contract for funds transfer. 
The network shows links between creators of different NFT projects involved in a rug pull. 
The illicit accounts in the figure are identified using `Fake phishing' and `Spam' labels from Etherscan's Label Word Cloud. 
The long trails of blue nodes represent chain of temporary accounts used to execute multiple transfers between multiple addresses of creators and illicit accounts.

\section{Patterns of distinct entities}\label{app:c}

\begin{figure}[H]
    \centering
    \includegraphics[scale=0.45]{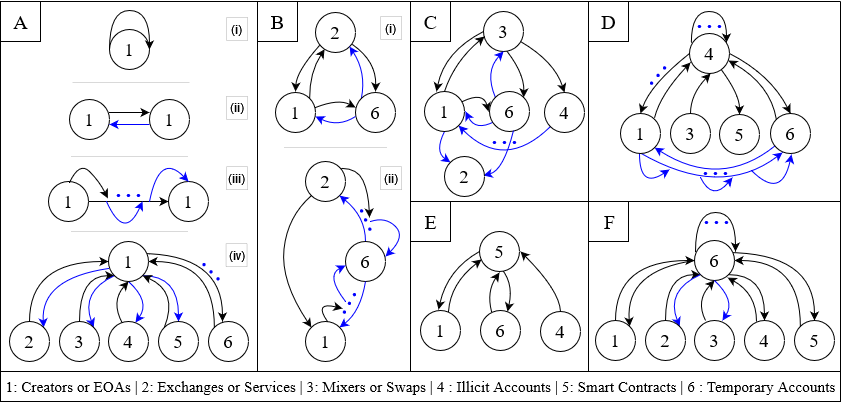}
    \caption{Interaction patterns of all entities with each other. Black arrows denote the first transaction whereas blue arrows denote the subsequent ones. Blue dots represent transactions from sender to the receiver through a chain of receiving entities.}
    \label{fig:allPatterns}
    \Description{All possible interactions from one entity to all others. This includes self-transactions, direct incoming and outgoing transfers and trail of transfers between any two entities.}
\end{figure}

Figure \ref{fig:allPatterns}(A) displays four behavioral patterns of a creator or wallet. Figure \ref{fig:allPatterns}(A.i) represents cases with the same sender and receiver addresses and transfers of 0 ETH (a symbol given to ether) between them. These transactions are an attempt to cancel or replace a pending transaction on Ethereum\cite{selfDirectedTxns}, which are filtered as they do not add value to this study. Figure \ref{fig:allPatterns}(A.ii) shows a direct exchange of money between creators and \ref{fig:allPatterns}(A.iii) shows a trail of transfers between multiple creators. Figure \ref{fig:allPatterns}(A.iv) displays relation between a creator and all other entities. 

Incoming funds to exchanges come either from creators or through their temporary accounts (cf. Figure \ref{fig:allPatterns}(B.i)). A creator funds a temporary account, which sends those funds to exchanges to mask direct trades. Figure \ref{fig:allPatterns}(B.ii) shows that a creator is funded by an exchange and transfers funds again to exchanges through multiple temporary accounts. Similarly, funds are sent to creators from an exchange through multiple temporary accounts. Creators and illicit accounts receive funds from mixers and transfer those funds to exchanges directly or using temporary accounts (cf. figure \ref{fig:allPatterns}(C)). 

Funds exchanged among multiple illicit accounts and creators are shown in \ref{fig:allPatterns}(D). It also shows that funds from several creators and their temporary addresses are accumulated in one illicit address. Illicit addresses also receive funds from mixers or swaps; and at times, transfer funds to contract addresses. Smart contracts transfer funds to creators or wallets as shown in \ref{fig:allPatterns}(E). Figure \ref{fig:allPatterns}(F) displays all possible interactions of temporary accounts, which are the most traversed addresses, with other entities. 

\begin{figure}[H]
    \centering
    \includegraphics[scale=0.2]{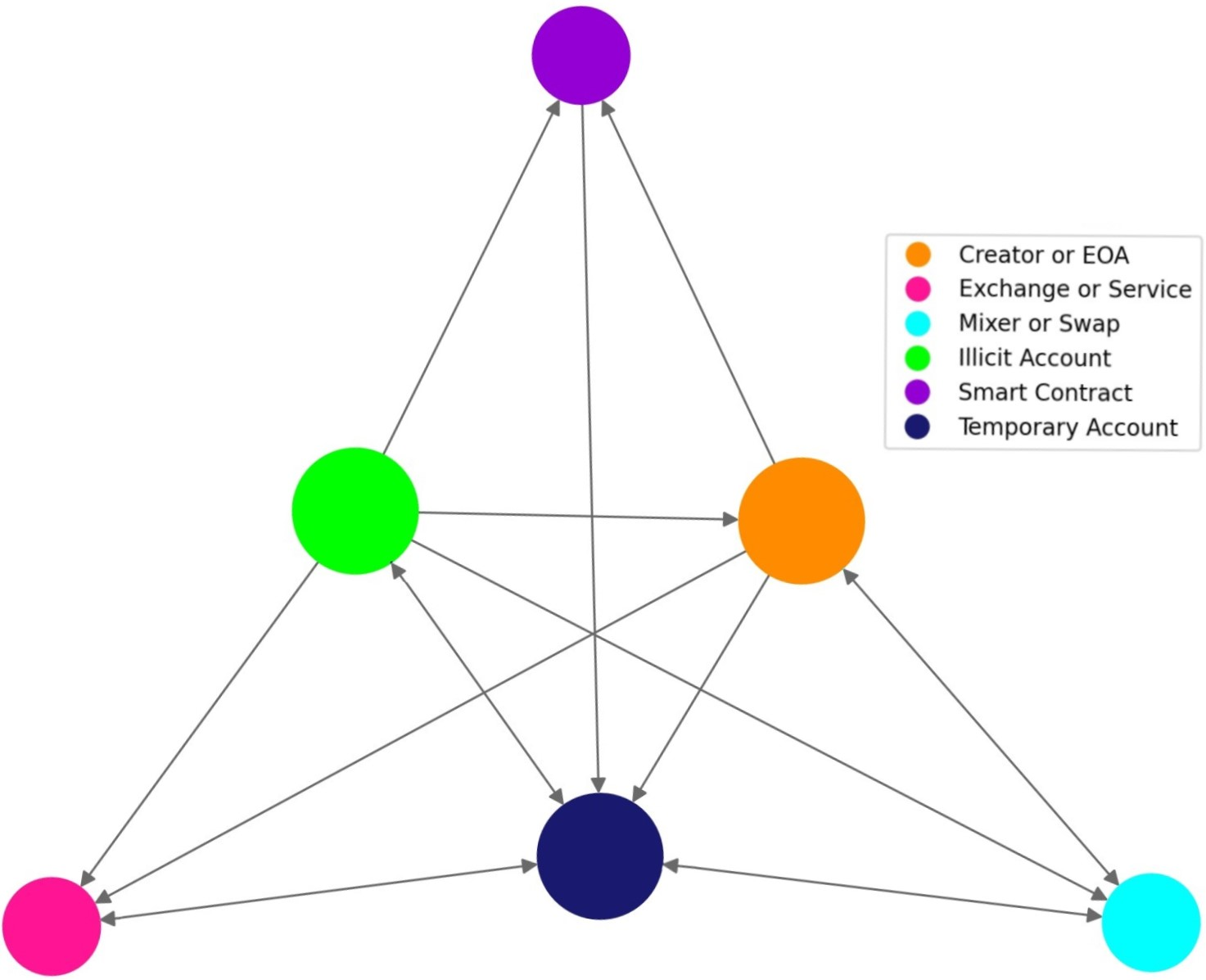}
    \caption{Interconnections of all entities with each other where degree of each node contributes to its size.}
    \label{fig:entitiesGraph}
    \Description{Connections between the six entities. Temporary account has the most number of incoming edges, followed by creators and illicit accounts.}
\end{figure}

Figure \ref{fig:entitiesGraph} shows the direct interlinks between the six entities that are considered in this study. Since the node size is according to their degree, nodes representing the creators, illicit accounts and temporary accounts become the central nodes.

\section{Flow of Funds}\label{app:d}

The volume of funds sent and received by the creators of rug-pulled projects totals approximately 384 million USD or 123,312 ETH. We determine that the volume received by these addresses is \$83.7 million USD. The majority of funds come through exchanges and services. Note that, we use the historical \textit{ether} price (the ether price at the time of the transaction) to convert \textit{ethers} to US dollars. 

\begin{figure}[H]
    \centering
    \includegraphics[scale=0.5]{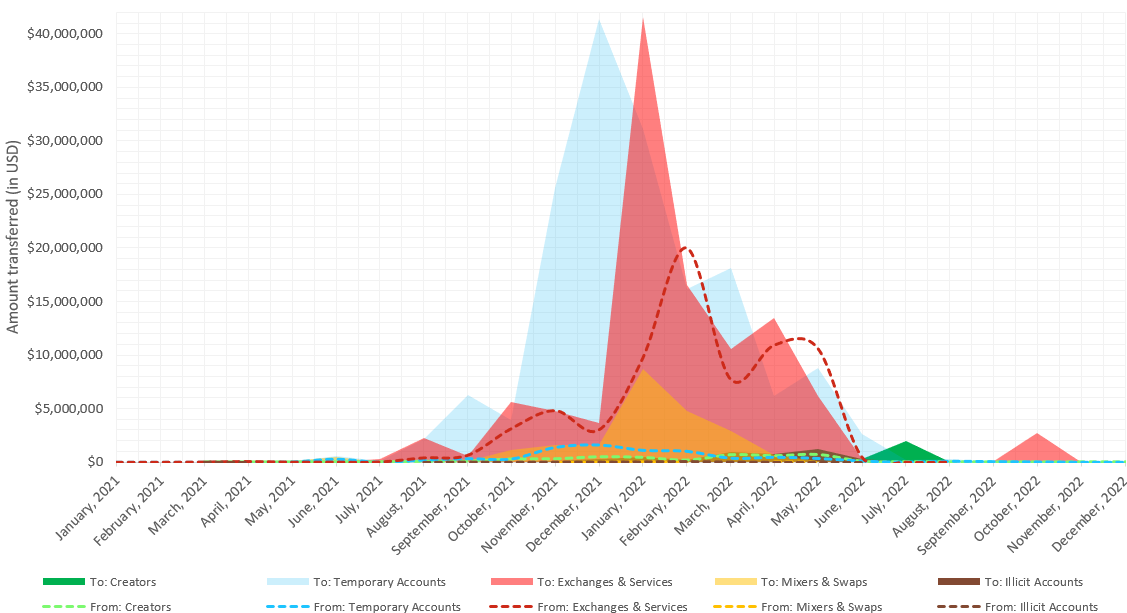}
    \caption{Flow of funds among a creator or creator's wallet (green), a temporary account (light blue), an exchange, marketplace, or service (light red), a mixing or swapping service (yellow), or any illicit account. The transferred amount is calculated month-wise. Dashed lines represent funds received from other entities, whereas colored area curves represent funds transferred to those entities.}
    \label{fig:8}
    \Description{Area chart showing flow of funds between creators and temporary accounts, exchanges, mixers, swaps and illicit accounts.}
\end{figure}

In this section, we elaborate the flow of funds between creators and other entities and list the most popular names of the involved entities along with their corresponding values. Figure \ref{fig:8} shows the incoming and outgoing movements of creator's funds from and to the temporary accounts, exchanges and services, mixers and swap services, and illicit accounts. The dashed blue line denotes the funds sent from  temporary accounts to creators. This depicts the funds exchanged between two creators via common temporary account(s). 
Funds represented by dashed blue line are clearly very low. On the other hand, the blue area curve representing outgoing funds from creators to temporary accounts is significantly high. 
The difference between incoming and outgoing funds indicates that the funds received by temporary accounts are transferred from one temporary account to the other, multiple times, before being sent to other entities. Our inference from this behavior is that, since these addresses are interrelated, they serve primarily to conceal the movement of funds from ordinary users and investigators.

As the number of rug pulls increased between October 2021 and March 2022, the red area curve indicate creators cashing out the investors’ money. 
Consider the red area curve (representing funds sent by creators to exchanges) between July 2021 and September 2021 as `a1' and the red dashed line (representing funds sent from exchanges to creators) rising from September 2021 as `a2'. These area curves show that creators traded funds from previous rug pulls, `a1', which returned profit from exchanges or in form of royalties, `a2'. This also indicates that creators use the received funds to attempt more rug pulls.

For each entity, there are significant differences between the dashed lines and colored area curves, representing the incoming and outgoing funds of creators, respectively. We expand our study to discover more insights on the flow of funds with respect to each entity.
 
\subsection{Exchanges or Services:} We discover 756 transfers in which the creators of NFTs received funds from an exchange or a service, totaling \$71.7 million USD. By analyzing the transaction data, we identify the exchanges or services that are used to transfer funds to the creators involved in rug pulls. The top 10 exchanges or services (based on funds volume) are listed in Table \ref{table:2}. 

\begin{table}[ht]
\begin{center}
\caption{Top 10 exchanges or services funding creators} 
\begin{tabular}{|l|l|c|} 
\toprule
\textbf{Funding Source} & \textbf{Type of Source} & \textbf{Amount (in USD)} \\ [0.5ex]
\midrule
\textbf{OpenSea} & Service & 65,912,490 \\
\textbf{Binance} & Exchange & 2,111,924 \\ 
\textbf{Gemini} & Exchange & 1,139,954 \\ 
\textbf{Crypto} & Exchange & 967,434 \\ 
\textbf{Coinbase} & Exchange & 834,934 \\ 
\textbf{KuCoin} & Exchange & 264,026 \\
\textbf{OKEx3} & Exchange & 129,841 \\ 
\textbf{Gate} & Exchange & 104,956 \\
\textbf{Kraken4} & Exchange & 66,994 \\ 
\textbf{Litebit} & Exchange & 61,718 \\
\bottomrule
\end{tabular}
\label{table:2} 
\end{center}
\end{table}

Here, OpenSea represents a payment service including royalties and other payments related to the NFT project. Apart from the top 10 listed in Table \ref{table:2}, nine creators received funds from FTX exchange\cite{ftx}. 
We discover 851 transfers in which the creators of NFTs transferred funds to exchanges, totaling \$108.5 million USD. 
The top 10 exchanges (based on highest amount) receiving funds from creators are listed in Table \ref{table:3}. Here, OpenSea represents OpenSea Wyvern exchange\cite{wyvern}.

\begin{table}[ht]
\begin{center}
\caption{Top 10 exchanges or services receiving funds from creators} 
\begin{tabular}{|l|l|c|} 
\toprule
\textbf{Receiving Entity} & \textbf{Type of Entity} & \textbf{Amount (in USD)} \\ [0.5ex]
\midrule
\textbf{BlockFi} & Exchange & 72,878,210 \\
\textbf{Binance} & Exchange & 7,863,101 \\ 
\textbf{OpenSea} & Exchange & 5,868,566 \\ 
\textbf{Coinbase} & Exchange & 5,174,934 \\ 
\textbf{OKEx3} & Exchange & 2,855,953 \\ 
\textbf{Gnosis} & Service & 1,870,136 \\ 
\textbf{Gemini} & Exchange & 1,559,814 \\ 
\textbf{KuCoin} & Exchange & 900,480 \\
\textbf{1inch} & Exchange & 491,963 \\
\textbf{Kraken} & Exchange & 455,408 \\
\bottomrule
\end{tabular}
\label{table:3} 
\end{center}
\end{table}

Consider the corresponding `amount' of first funding source (cf. Table \ref{table:2}), OpenSea, as `f1' and that of first receiving entity (cf. Table \ref{table:3}), BlockFi, as `f2'. We make following assertions:
(i) By comparing `f1' and `f2' with the `amount' of other entities in both tables, it is evident that `f1' and `f2' are significantly higher than all the others. 
(ii) Knowing that the total funds received by creators are \$71.7 million USD, we conclude that 92\% of initial funds are sent from OpenSea. 
From (i) and (ii), we infer that (iii) `f1' is a part/subset of `f2'.
As per our dataset from January to May 2022, we demonstrate the above inference from the creator's transactions of a project named `Azuki':
(i) The creator received a total of 43,137,590 USD from OpenSea in form of royalties and payments. The transfers occurred in batches and 4 - 12 times every month.
(ii) Creator transferred 72,878,210 USD to BlockFi, in batches. The transfers occurred 2-5 times every month.
This shows a direct flow of funds from one exchange to an NFT project creator and further to another exchange. Note that, the creator in context interacted (received and transferred funds) with other exchanges as well, however, BlockFi received the highest number of funds among all other exchanges. 
It is important to observe that the highest amount of funds received by an exchange is from a single creator address which received most of the initial funding from OpenSea royalties. Since, royalties come from sales of an NFT project, initial funds received from royalties indicate the presence of one or more previous projects related to `Azuki'. Involvement of `Azuki' in rug pull further indicates the malicious intentions behind creation of other related projects. Here, we seek to demonstrate the impact of a single rug pulled project on the financial loss faced by investors.
Note that, the top 10 exchanges, based on total funds volume, also represent them as the most popular exchanges used by creators involved in rug pulls.


\subsection{Creators:} Around \$3.7 million USD changed hands among numerous creators via direct transfers. We identify the creators who received the highest amount of funds from other creators involved in rug pulls. The information on identified creators with their corresponding funds is provided in Table \ref{table:4}. 
\begin{table}[ht]
\begin{center}
\caption{Top 10 creator addresses receiving funds from creators} 
\begin{tabular}{|l|l|c|} 
\toprule
\textbf{Sending Creator} & \textbf{Receiving Creator} & \textbf{Amount (in USD)} \\ [0.5ex]
\midrule
MoonApeLab & \textbf{MoonApeLabLoot} & 520,041 \\
HighriseLand & \textbf{HighriseLand}\_ & 394,950 \\ 
MoonApeLabLoot & \textbf{MoonApeLab} & 387,084 \\ 
AcidDragons & \textbf{Pals} & 162,181 \\ 
PartyApeBillionaireClub & \textbf{IceWorld} & 158,829 \\ 
Mutts & \textbf{BitFrenchie} & 151,190 \\ 
Gunslingers & \textbf{Pals} & 104,676 \\ 
RocketBunny & \textbf{ApocalypticApes} & 101,060 \\
Pals & \textbf{AcidDragons} & 97,884 \\
NobleKnights & \textbf{ShaolinSamurai} & 95,836 \\
\bottomrule
\end{tabular}
\label{table:4} 
\end{center}
\end{table}
The highest amount of funds transferred directly between two creators is done by creators of 'Moon Ape Lab' and 'Moon Ape Lab Loot' NFT projects. Both projects are associated with the `Rug pull mafia' group.
By identifying the creators who are most active in funding other creators involved in rug pulls, we also identify the rug pull groups that those creators belong to. We find that the three most active rug pull groups are `Rug pull mafia', `Mutant Tiny Dinos' and `Business Ape Club'.

Table \ref{table:5} displays funds received by creators or wallets of creators. Out of \$3.7 million USD exchanged between creators, we observe that around 1.1 million USD are transferred to wallets of creators.

\begin{table}[ht]
\begin{center}
\caption{Top 10 creator addresses sending funds to creators or their related wallets} 
\begin{tabular}{|l|l|c|} 
\toprule
\textbf{Sending Creator} & \textbf{Receiving Creator} & \textbf{Amount (in USD)} \\ [0.5ex]
\midrule
\textbf{TheCompanion} & BluttlesWallet & 695,059 \\
\textbf{MoonApeLab} & MoonApeLabLoot & 520,041 \\ 
\textbf{HighriseLand} & HighriseLand\_ & 394,949 \\ 
\textbf{MoonApeLabLoot} & MoonApeLab & 387,084 \\ 
\textbf{AcidDragons} & Pals & 162,181 \\ 
\textbf{PartyApeBillionaireClub} & IceWorld & 158,829 \\ 
\textbf{Mutts} & BitFrenchie & 151,190 \\ 
\textbf{Skulltoons} & SkulltoonsWallet & 132,834 \\
\textbf{Fuji} & FujiWallet & 119,734 \\
\textbf{Gunslingers} & Pals & 104,676 \\
\bottomrule
\end{tabular}
\label{table:5} 
\end{center}
\end{table}

17 creators are found to send funds directly to themselves, with the highest amount transferred being 27,494 USD by the creators of 'Apocalyptic Apes' project. 
Here, there are two assertions: (i) the receiving and sending entities are same, i. e., creators and (ii) creators and their connected wallets belong to the same entity, i. e., creators (cf. Appendix Section~\ref{app:A} \ref{app:b}).
Considering (i) and (ii), we show that though creators and wallets belong to the same entity, to understand their distinct flow of funds, we keep creator wallets as a separate entity. Thus, we get more information on flow of funds by observing that creator wallets are only the receiving entity, unlike creators who are both senders as well as receivers. 
Keeping wallets distinct, clarifies our definition of creator's wallets as the addresses that receive and store profits of creators. Note that, at the time of recording our dataset, we find wallets storing funds of creator, however, wallets are capable of acting as any other entity (depending on the transaction behavior) later.


\subsection{Illicit accounts:} As per our dataset, there are 61 illicit addresses. Approximately, \$193,381 USD are obtained from 40 illegitimate accounts by 76 distinct creator addresses. One illicit address sent funds to 20 creators of the `Rug pull mafia' group. 19 creators transferred 18,306 USD to 22 illicit addresses.

\subsection{Mixers and Swap services}
\$46,871 USD are obtained through swapping and mixing services. We segregate mixers and swap services and discuss their individual inferences below. Since we only have one mixer service in our dataset (i. e., Tornado Cash), we display the important inferences related to funds flow between Tornado Cash and creators.

\subsubsection{Tornado Cash:} We identify three addresses related to Tornado Cash in our dataset. A total of \$7,986 USD is transferred from Tornado Cash to three creators, belonging to three different rug pull groups: `Business Ape Club', `Rug pull mafia' and `Ivy Girls'. 

Project named `Not A Secret Project' received \$3,064 USD, making it the top recipient of funds from Tornado Cash. Moreover, 25 creators, belonging to 9 rug pull groups, laundered around \$7.85 million USD through Tornado Cash. The majority of creators belong to `Rug pull mafia' group, transferring \$602,498 USD to Tornado Cash. However, the highest amount of funds, \$4.25 million USD (55\% of the total funds sent to Tornado Cash), are transferred from the creator of 'Undead Pastel Club' project. Here again, a single creator contributed to the majority of funds towards a sanctioned mixer. 

\subsubsection{Swap services:} In our dataset, ChangeNOW\cite{changenow} and Uniswap\cite{uniswap} are the leading swap services that funded creators. Six creators received \$38,886 million USD from swap services. Table \ref{table:6} shows that creators received \$24,631 USD from ChangeNOW and \$14,254 USD from Uniswap. 88 distinct creators transferred around \$14.5 million USD to swap services. The top swap services used by creators involved in rug pulls to transfer their accumulated funds are: `Metamask Swaps'\cite{metamaskswap}, `Uniswap' and `Sushi swap'\cite{sushiswap}. 

\begin{table}[ht]
\begin{center}
\caption{Swap services sending funds to creators} 
\begin{tabular}{|l|l|c|} 
\toprule
\textbf{Swap Service} & \textbf{Creator} & \textbf{Amount (in USD)} \\ [0.5ex]
\midrule
\textbf{ChangeNOW} & MoonApeLab & 13,110 \\
\textbf{ChangeNOW} & DerpyApes & 10,434 \\ 
\textbf{Uniswap} & BusinessApeClub & 7,124 \\ 
\textbf{Uniswap} & ShaolinSamurai & 7,110 \\ 
\textbf{ChangeNOW} & DerpyApeMfers & 1,087 \\ 
\textbf{Uniswap} & MutantAlienApeYachtClub & 20 \\
\bottomrule
\end{tabular}
\label{table:6} 
\end{center}
\end{table}

Table \ref{table:7} shows that most of the funds are transferred to MetaMask swap service, receiving around \$7.2 million USD. Creators transferred \$4.6 million USD through Uniswap V3 and \$1.8 million USD through Uniswap V2, respectively. In total, Uniswap received \$6.4 million USD from creators.

\begin{table}[ht]
\begin{center}
\caption{Top 10 creator addresses sending funds to swap services} 
\begin{tabular}{|l|l|c|} 
\toprule
\textbf{Creator} & \textbf{Swap Service} & \textbf{Amount (in USD)} \\ [0.5ex]
\midrule
\textbf{BoredBunny} & MetamaskSwap & 3,157,642 \\
\textbf{SuperFatApes} & MetamaskSwap & 1,646,565 \\ 
\textbf{LilHippo} & UniswapV3 & 1,122,690 \\ 
\textbf{BabyElonWorld} & UniswapV3 & 1,048,886 \\ 
\textbf{Pals} & MetamaskSwap & 923,782 \\ 
\textbf{BabyElonWorld} & MetamaskSwap & 690,668 \\ 
\textbf{ShaolinSamurai} & UniswapV2 & 505,774 \\ 
\textbf{ZombieMonkeys} & UniswapV3 & 415,674 \\
\textbf{DegenApeClub} & UniswapV2 & 314,280 \\
\textbf{ShaolinSamurai} & UniswapV3 & 313,294 \\
\bottomrule
\end{tabular}
\label{table:7} 
\end{center}
\end{table}

The creator of project `Apocalyptic Apes' transferred \$197,349 USD to 5 different swap services, which is the highest number of swap services used by one creator. The top 3 creators who sent funds to more than one swap service is shown in Table \ref{table:8}.

\begin{table}[ht]
\begin{center}
\caption{Top 3 creators using multiple swap services} 
\begin{tabular}{|l|l|c|} 
\toprule
\textbf{Creator} & \textbf{Swap Service} & \textbf{Amount (in USD)} \\ [0.5ex]
\midrule
\textbf{ApocalypticApes} & UniswapV2 & 167,968 \\
& UniswapV3 & 23,681 \\
& GenieSwap & 4,152 \\
& PlasmaSwap & 732 \\
& MetamaskSwap & 629 \\
& GemSwap & 187\\ [0.5ex]
\midrule
\textbf{ShaolinSamurai} & UniswapV2 & 505,774 \\
& UniswapV3 & 336,811 \\
& SushiSwap & 23,517 \\
& GemSwap & 618 \\ [0.5ex]
\midrule
\textbf{WarriorApes} & UniswapV2 & 5,952 \\
& MetamaskSwap & 1,665 \\
& UniswapSNX6 & 0.69 \\
& SynthetixOldDepot & 0.13 \\
& UniswapOldsUSD & 0.03 \\ [0.5ex]
\bottomrule
\end{tabular}
\label{table:8}
\end{center}
\end{table}


\subsection{Temporary accounts:} With 2,601 transfers, the migration of funds from various temporary accounts to creators totals \$7.9 million USD. We observe from transaction behaviors of 8 NFT projects that creators transfer funds through 15–20 temporary accounts, creating a trail which ends up at a wallet address. In one case, there is a trail of 60 temporary accounts. Another behavior we observe with the temporary accounts is that the transfers of funds occurs in smaller portions or batches.    
Creators transferred \$80.9 million USD to temporary accounts. 
We find that 365 creators used temporary accounts to transfer funds to other entities. Creator of project `FatApeClub' is identified to transfer \$13.1 million USD, the highest amount of funds sent to temporary accounts. `IvyGirls', a rug pull group, of 38 NFT projects used one temporary account to transfer funds between the creators of the group. It is observed that the temporary account received funds from 8 distinct creators and sent funds to other 4. There are four more creators of the same group who exchanged funds with each other repeatedly. Another temporary account sent funds to 13 distinct creators, belonging to 4 rug pull groups. 
This association of temporary accounts and rug pull groups is also discussed in Section \ref{ssubsec:rpgroups}.

\end{document}